\documentclass[aps,pre,nofootinbib,showpacs,twocolumn]{revtex4-1}%
\usepackage{amssymb,latexsym,mathrsfs}
\usepackage{amsfonts}
\usepackage{mathrsfs}
\usepackage{amsmath}
\usepackage{graphicx}
\usepackage{color}

\newcommand{\beq}{\begin{equation}}
\newcommand{\eeq}{\end{equation}}
\newcommand{\beqn}{\begin{eqnarray}}
\newcommand{\eeqn}{\end{eqnarray}}
\newcommand{\bearr}{\begin{array}}
\newcommand{\enarr}{\end{array}}

\newcommand{\eps}{\epsilon}

\def\bea{\begin{eqnarray}}
\def\eea{\end{eqnarray}}
\def\ba{\begin{array}}
\def\ea{\end{array}}

\begin{document}
\title{Spin diffusion in one-dimensional classical Heisenberg model} 
\author{Debarshee  Bagchi}
\email[E-mail address: ]{debarshee.bagchi@saha.ac.in}
\affiliation{Theoretical Condensed Matter Physics Division,\\ Saha Institute of Nuclear Physics,\\
1/AF Bidhan Nagar, Kolkata 700064, India.}
\date{\today}

\begin{abstract}
The problem of spin diffusion is studied numerically in one-dimensional classical
Heisenberg model using a deterministic odd even spin precession dynamics.
We demonstrate that spin diffusion in this model, like energy diffusion, is normal
and one obtains a long time diffusive tail in the decay of
autocorrelation function (ACF). Some variations of the model with different coupling schemes
and with anisotropy are also studied and we find normal
diffusion in all of them. A systematic finite size analysis of the Heisenberg model also 
suggests diffusive spreading of fluctuation, contrary to previous claims
of anomalous diffusion.

\end{abstract}
\maketitle

The classical Heisenberg model \cite{fisher,joyce} has been extensively studied, both
analytically and numerically, for several decades and has become a
prototypical model for magnetic insulators. However, one important question
that still awaits a conclusive answer is regarding the time dependent behavior
of the spins, particularly at very high temperature.
In the hydrodynamic limit, the dominant mode of fluctuation
spreading in this system is believed to obey of the standard diffusion phenomenology.
In absence of any microscopic theoretical formalism, studies were mostly
numerical, generally involving calculation of time correlation functions.
Although the phenomenology of spin diffusion is an old concept \cite{old, hove},
its validity in classical Heisenberg model has been vigourously debated
in recent times. 
Although much effort \cite{muller, landau, mct, nonlinear, other, error, 
breakdown, breakdown-comment} has been devoted to understand whether spin
diffusion in this system is normal or anomalous, a convincing conclusion
is yet to be reached.
Settling this question is not only conceptually important e.g., in 
understanding transport properties of spin systems, but also has
direct implications in routinely performed experiments e.g., NMR 
and ESR in magnetic compounds \cite{expt-1,expt-2,expt-3,expt-4,expt-5}.
In the following, we present a brief outline of the diffusion phenomenology
and review some of the earlier studies in this direction.

Let us consider a one-dimensional chain containing Heisenberg spins $\{\vec{S}_i\}$
(three dimensional unit vectors) where, $i = 1, 2,\dots, N$ with periodic
boundary conditions, i.e., $\vec{S}_{N+1} \equiv \vec{S}_1 $. The 
Hamiltonian is given by,
\begin{equation}
\mathcal{H} = - \sum_{i=1}^{N} K_i ~ \vec{ S}_i\cdot \vec{ S}_{i+1},
\end{equation}
where $K_i$ is the interaction strength between the spins $\vec S_i$
and $\vec S_{i+1}$; the spin-spin coupling is ferromagnetic for $K_i > 0$
and anti-ferromagnetic if $K_i < 0$.
The microscopic equation of motion can be written as,
\begin{equation}
\frac d{dt} {\vec{ S}_i} = \vec{S}_i \times \vec{B}_i,
\label{eom}
\end{equation}
where $\vec{B}_i = K_{i-1} \vec{S}_{i-1} + K_i \vec{S}_{i+1}$ is the local 
molecular field experienced by the spin at site $i$.  Clearly, 
Eq. (\ref{eom}) conserves 
(i) the total energy $E = \sum_i E_i = - \sum_i  K_i ~\vec{ S}_i\cdot \vec{ S}_{i+1}$, and
(ii) the total spin $\vec S = \sum_i \vec{S}_i$.

Since there is no long range order in this system at any finite 
temperature and because of the conservation of total spin, 
%$\vec{S}_T = \sum_i \vec{S}_i$,
the spin fluctuation in the hydrodynamic limit is expected to
follow a continuity (diffusion) equation
% 
% 
% 
%\begin{equation}
${\partial}_t \vec{S}_q(t) = - D_s q^2 \vec{S}_q(t)$,
%\label{Sdiff}
%\end{equation}
% 
% 
% 
where $\vec{S}_q(t)$ is the (discrete) Fourier transform of $S_i(t)$ and
$D_s$ is the spin diffusion constant. % \in [0,1]$.
A similar equation holds
for the energy density (since total energy is also a constant of motion).
%
% %
% \begin{equation}
% \frac{\partial}{\partial t} E_q(t) = - D_e q^2 E_q(t),
% \label{Ediff}
% \end{equation}
%
%
% $D_e$ being the energy diffusion coefficient. 
%Eq. (\ref{Sdiff}) 
The continuity equation implies that 
in the hydrodynamic limit (small $q$ and large $t$) the spin-spin 
correlation function $A_s(\vec{q},t) \equiv \langle \vec{S}_q(t)
\cdot \vec{S}_{-q}(0) \rangle$ decays with time exponentially, i.e.,
% 
% 
% 
%\begin{equation}
$ A_s(\vec{q},t) \sim e^{-D_s q^2 t}.$
%\end{equation}
A direct consequence of this is that the spin autocorrelation function (ACF)
$A_s(t) \equiv \frac 1N \sum_i \langle \vec{S}_i(t) \cdot \vec{S}_i(0) \rangle$ at late times
decays with a power law tail
%\begin{equation}
$A_s(t) \sim t^{-\alpha}$.
%\end{equation}
As predicted by the diffusion phenomenology, 
the exponent $\alpha$ is equal to $1/2$ in one dimension.
This is also true for energy 
ACF $A_e(t) \equiv \frac 1N \sum_i  \langle E_i(t)E_i(0) \rangle \sim t^{-1/2}$.

Recently, the problem of spin diffusion in this model was studied by M\"{u}ller \cite{muller}
and it was reported that $\alpha = 0.609 \pm 0.005$ (1D) in the
hydrodynamic limit, thus significantly differing from the spin diffusion prediction.
% Gerling
Following this, Gerling \textit{et. al.} \cite{landau}
performed extensive numerical studies with larger system sizes and
for longer times. They strongly opposed the claim made in Ref. \cite{muller}
and demonstrated that the slope of the ACF
slowly decreases as $t$ is increased. Nevertheless, they suggested
that the problem is computationally difficult since the non-asymptotic
behaviour of the spin ACF is quite pronounced.
% Error - Liu
In yet another work \cite{error}, it was concluded that in numerical simulation it
is not possible to observe the $t^{-1/2}$ behavior, even if it exists,
due to the fact that the numerical scheme introduces computational errors
and this violated the conservation of total spin $\vec S$. The error propagation
affects the decay of the ACF and makes it anomalous.
It was suggested that the correlation function may show a 
crossover from non-diffusive to diffusive behavior and the characteristic 
crossover time will depend on the precision of the numerical scheme employed.

% Breakdown - Bonfim and Reiter
Another subsequent numerical work in this direction \cite{breakdown}
however claimed that although energy diffusion is normal, spin diffusion
has an anomalous behavior. 
\cite{breakdown-comment}.
A coupled-mode theory of spin fluctuation \cite{mct}
suggested that spin diffusion is anomalous with $A_s(t) \sim t^{-2/5}$ 
asymptotically. 
Authors in Ref. \cite{other} studied few variants of the Heisenberg 
models (alternate coupling, random coupling etc.) and suggested that 
spin diffusion is probably normal with alternate coupling but is
anomalous with random coupling.
A nonlinear dynamics study \cite{nonlinear} presented numerical results in
support of anomalous diffusion and computed $z \approx 1.67$, which implies $\alpha
 = z^{-1} \approx 0.6$. %from finite size analysis of the time correlation function for different system size. 
Thus, most of the previous works refute the
validity of normal spin diffusion in this system.

%Our work
In this work, we re-investigate spin diffusion in classical Heisenberg
model on a ring, using a discrete time odd even
dynamics (DTOE) \cite{our}. % to evolve the system with time.
We compute the temporal decay of the energy and spin ACFs, and its cumulative average (defined later). 
We perform extensive simulation on large system sizes ($N = 5000 - 20000$)
and for very large times ($t = 10^6$). To the best of our knowledge,
such large scale simulation has never been performed in this system to 
study spin diffusion.
We demonstrate convincingly that spin diffusion is normal and the exponent 
$\alpha = 1/2$ (within numerical accuracy).
We also study a few other variants of the usual Heisenberg model and show that
diffusion process is also normal in those cases. For small system sizes, 
the ACFs saturate after a characteristic timescale that depends on the system size.
We have performed systematic finite size analysis for small systems
which again indicates that spin diffusion is indeed normal.
In the following, we describe our numerical scheme and present the results.

% 
% \section{Numerical sumulation}
% \label{num-sim}
% 
% \subsection{DTOE dynamics}
% \label{dtoe}
\paragraph{DTOE dynamics:} 
% To study the model we use a scheme which we refer to as the
% discrete time odd even (DTOE) dynamics. 
This method of integrating the discretised version of equation of motion has been discussed
in detail elsewhere (see Ref. \cite{our}). 
For the sake of completeness, we present an outline of the DTOE dynamics here.

To integrate the equation of motion numerically, one would naively consider
a finite difference equation of the Euler form
\begin{equation}
\vec{ S}_{i,t+1} = \vec{S}_{i,t} + \Delta t \,\, \left[\vec{S} \times \vec{B}\right]_{i,t}
\label{f_diff}
\end{equation}
and update \textit{all} the spins at time $t$ and obtain their values
at the next time-step $t+1$. However, it can be shown that using
Eq. (\ref{f_diff}) directly and updating all the spins simultaneously,
lead to the violation of the conservation laws stated above for all
$\Delta t > 0$; the length of the spins is also not held constant \cite{our}.
A way to naturally preserve the length of spins $|\vec S_i|$ is by using
an alternative spin precession update equation (instead of Eq. \ref{f_diff})
\begin{equation}
\vec{S}_{i,t+1} = \left[\vec{S} \cos \phi + (\vec{S} \times \hat{B}) 
\sin \phi + (\vec{S}\cdotp\hat{B})\hat{B}(1-\cos \phi) \right]_{i,t},
\label{precess}
\end{equation}
where $\hat{B}_i = \vec{B}_i/|\vec{B}_i|$ and $\phi_i = |\vec{B}_i| \Delta t$ \cite{goldstein}.
%instead of  Eq. (\ref{f_diff}). 
This will, however, still violate
the conservation of total energy and total spin.

For the conservation of total energy, we use an odd-even spin update rule where
the dynamics described in Eq. (\ref{precess}) is numerically implemented by alternate parallel
updates of the spins on odd and even sublattices.
Thus, at each step, first, only even spins are 
updated using the spin precession dynamics Eq. (\ref{precess}) 
while the odd spins  are kept unaltered. Next, the spins on the odd sublattice 
are similarly updated.
It is straight forward to check  that update  of  any spin $\vec{S}_i$  affects  
only the  energy of  the neighbouring bonds $\eps_{i-1}$   and $\eps_i$, but their 
sum ($\eps_{i-1} + \eps_i$) remains constant.  Thus DTOE dynamics conserves energy
strictly and also naturally maintains the individual spin lengths.

However, the total spin $\vec S$ does not remain conserved using DTOE dynamics.
This is a general problem with any standard integration scheme;
the conservations are only approximately maintained depending on the
accuracy of the scheme.

\begin{figure}[htb]
\centerline
{
\includegraphics[width=3.5cm,angle=-90]{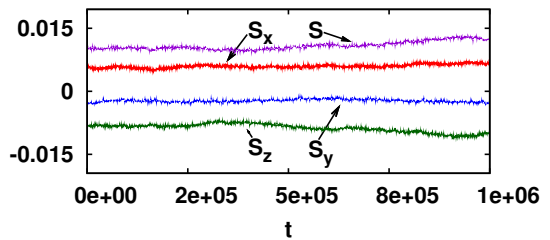}
}
\caption{
(Color online) Typical evolution of the magnitude of the total spin 
$S \equiv |\vec S|$ and its components $S_x, S_y, S_z$ using DTOE
dynamics with $N = 5000$, $\Delta t = 0.05$ and with uniform
coupling $K_i = 1$. All the four quantities are scaled by a
factor of $1/N$.
}
\label{fig:Scons}
\end{figure}

We will show that DTOE dynamics is still a better numerical scheme compared to other
conventional schemes (e.g. Euler method, Runge-Kutta method) since the accuracy here is naturally
higher for any given value of $\Delta t$ (as $E$ and $|\vec S_i|$ remain accurately
conserved for any arbitrary $\Delta t$). Thus, one can choose a relatively larger $\Delta t$ without
accumulating large numerical errors and thus probe the time dependent behavior
of the system at very late times.
Although choosing a larger $\Delta t$ essentially converts the equation of motion to 
a map, it can be shown that independent of the value of $\Delta t$, this dynamics
allows the system to settle to the correct equilibrium state \cite{our} and one
obtains the correct static spin correlations. Also, we have verified it thoroughly
that the asymptotic behavior of the ACFs remains unaltered
with smaller $\Delta t$, only the computation becomes more time consuming.

The time evolution of the total
spin and its components, from typical run using the DTOE dynamics, is shown in
Fig. \ref{fig:Scons}. We find that up to large times ($t \sim 10^6$), the total spin conservation
is approximately maintained; the magnitude $S$ and the components $S_x, S_y, S_z$
do not show any trend of an overall growth (or decay) with time. An Euler-like scheme
for the same values of the parameters $N$ and $\Delta t$ will however develop numerical instabilities
and `blow up' much before $t \sim 10^6$.
Thus, using DTOE dynamics, one can perform extensive numerical simulation of the system
and reliably determine the behavior of the system in the hydrodynamic limit. 
Below, we mention the details of our numerical simulation and present the results.

\paragraph{Results:} First, we simulate the case for which all interactions are positive (ferromagnetic)
and uniform i.e., $K_{i} = K$ (and set to unity without the loss of generality).
Starting from a random initial spin configuration, we evolve the system using
DTOE dynamics. We compute the ACFs $A_e(t)$ and $A_s(t)$, and study
their late time power law decay. However, the ACF has a slow
convergence to its asymptotic behavior. Again, in course of its decay with time,
the ACF either saturates if the system size is small or,
if the system size is large enough, the ACF continues
to decay and its the numerical value keeps decreasing. 
As such, it becomes computationally more and more
expensive to get well averaged data at late times. 
To circumvent these problems, instead of the the ACF,
we compute its cumulative average, defined as
\begin{equation}
C_s(t) = \frac 1t \int_0^t A_s(\tau) d \tau,
\label{cumu}
\end{equation}
where
%\begin{equation}
$A_s(t) =   \frac 1N \sum_{i=1}^N \langle \vec S_i(0) \cdot \vec S_i(t) \rangle$.
%\label{cumu}
%\end{equation}
%
% 
This cumulative autocorrelation function (CACF) $C_s(t)$ has the same asymptotic time dependence
as the ACF and therefore at late times $C_s(t) \sim t^{-1/2}$, if spin diffusion is normal. 
Moreover, this has the added advantage that the data for $C_{s}(t)$ is 
much less noisy than that of $A_{s}(t)$ and therefore its asymptotic time 
dependence can be computed with high accuracy. Likewise, one can define a
cumulative average $C_e(t)$ for $A_e(t)$.

\begin{figure}[htb]
\centerline
{
\includegraphics[width=3.45cm,angle=-90]{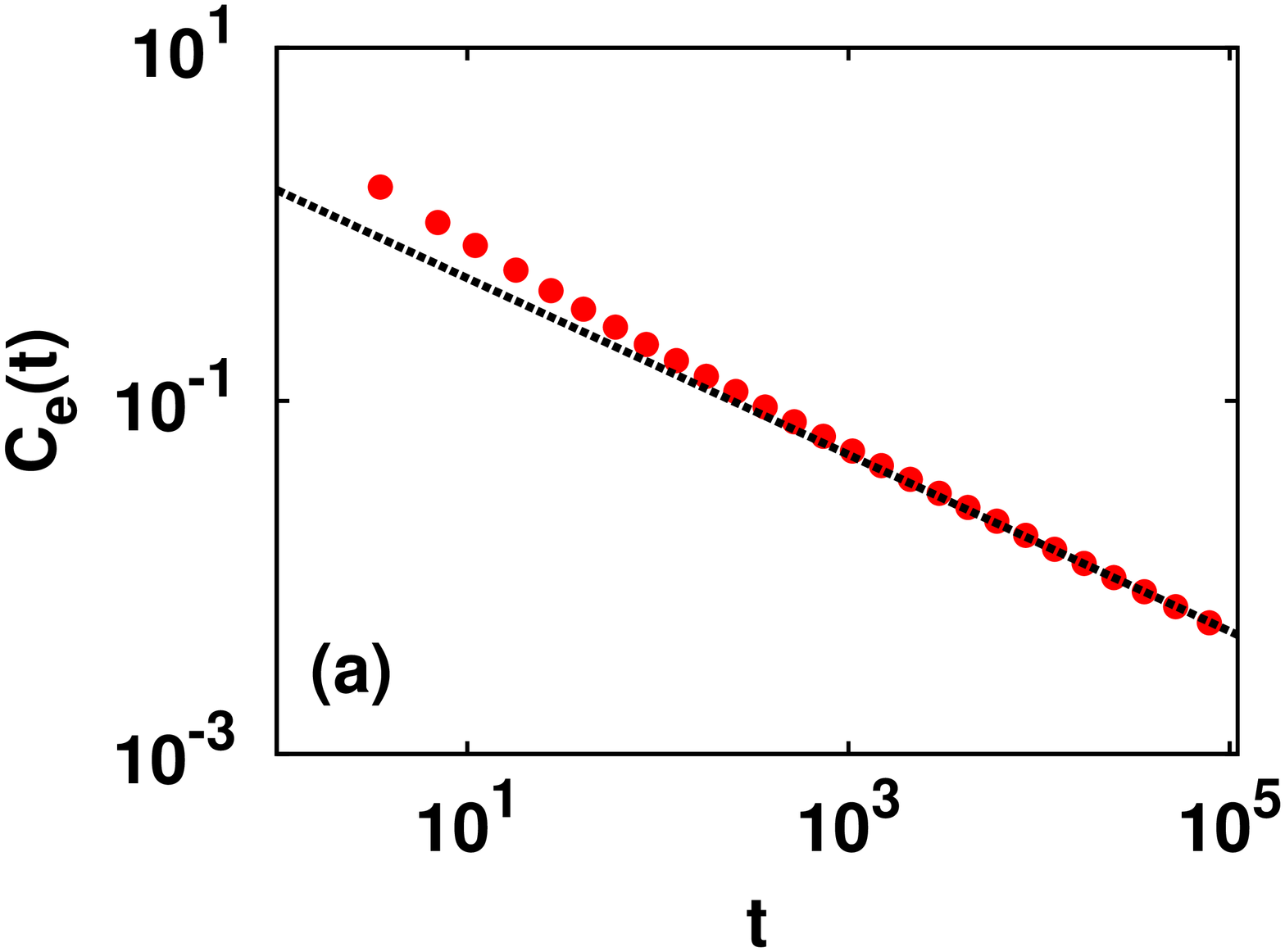}\hfill
\hskip-0.5cm
\includegraphics[width=3.45cm,angle=-90]{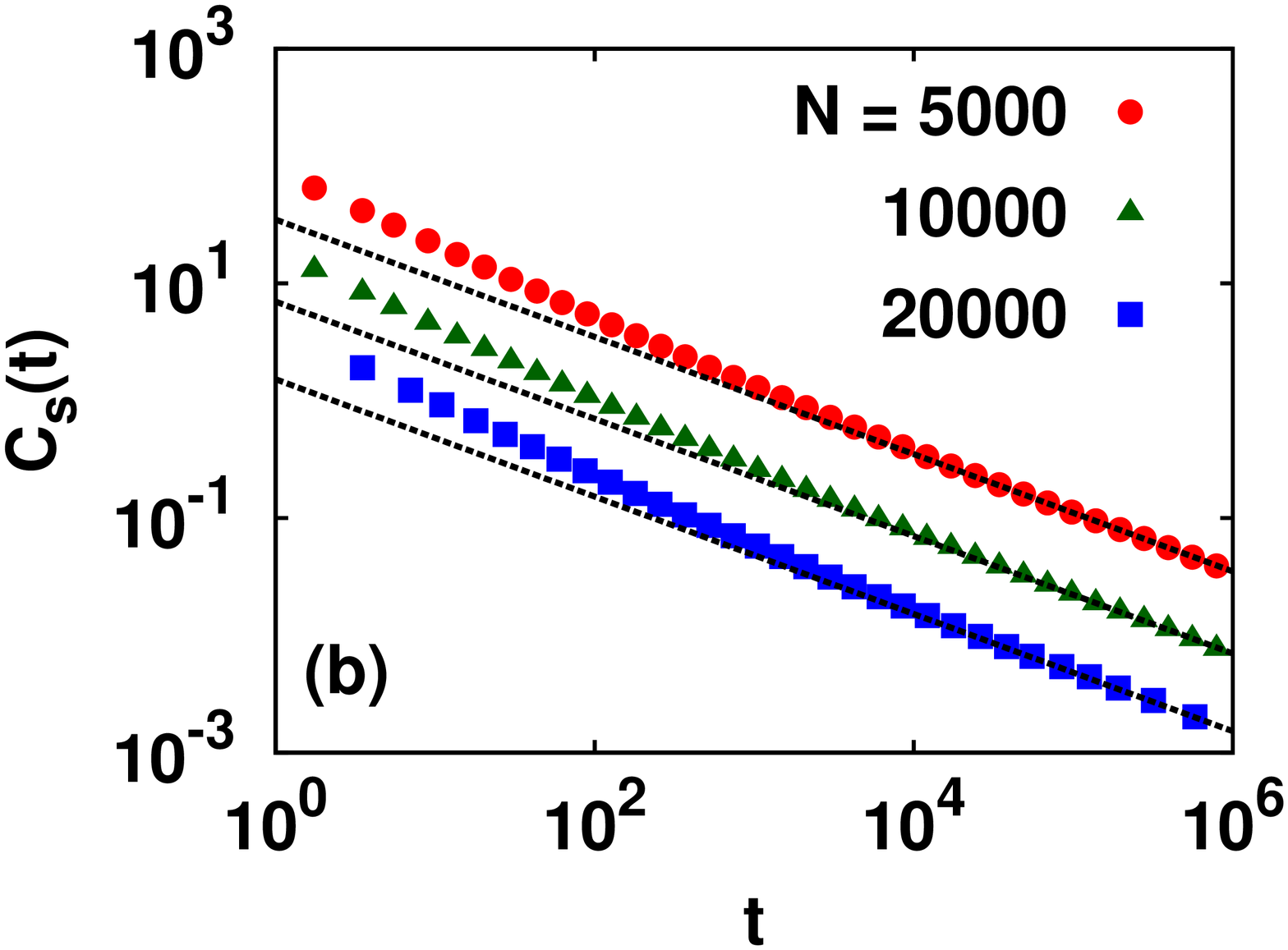}
}
\caption{
(Color online) Log-log plot of the autocorrelation functions for the uniform case (a) $C_e(t)$ 
for $N,\Delta t = 20000, 0.1$ averaged over $100$ independent realisations.
(b) $C_s(t)$ for $N,\Delta t = (5000,0.05),(10000,0.05)$ and $(20000,0.1)$, and averaged over
 $200, 100$, and $25$ independent realisations respectively. The data points are shifted along the $y$-axis
for better visibility. The broken lines have a slope $-0.5$
}
\label{fig:Uni-S}
\end{figure}

The functions $C_e(t)$ and $C_s(t)$ obtained using DTOE dynamics
are shown in Fig. \ref{fig:Uni-S}. 
At late times ($\sim t > 10^4$), we find that all the curves show a clear convergence 
to $t^{-1/2}$ (broken lines in the figures). 
The $t^{-1/2}$ decay of the energy CACF (Fig. \ref{fig:Uni-S}a) does not come as a surprise since
energy diffusion in this model was already known to be normal.
It has also been recently shown that energy transport in this model obeys Fourier's law \cite{savin,our}
for any nonzero temperature.
However, the decay of the spin CACF (Fig. \ref{fig:Uni-S}b) with an exponent $\alpha = 1/2$ for almost
two decades ($\sim 10^4 - 10^6$) is quite interesting.
In fact, this clearly indicates that, contrary to previous claims, spin diffusion is normal in the classical Heisenberg spin
system and spin fluctuation spreads diffusively.

\begin{figure}[]
\centerline
{
\includegraphics[width=3.5cm,angle=-90]{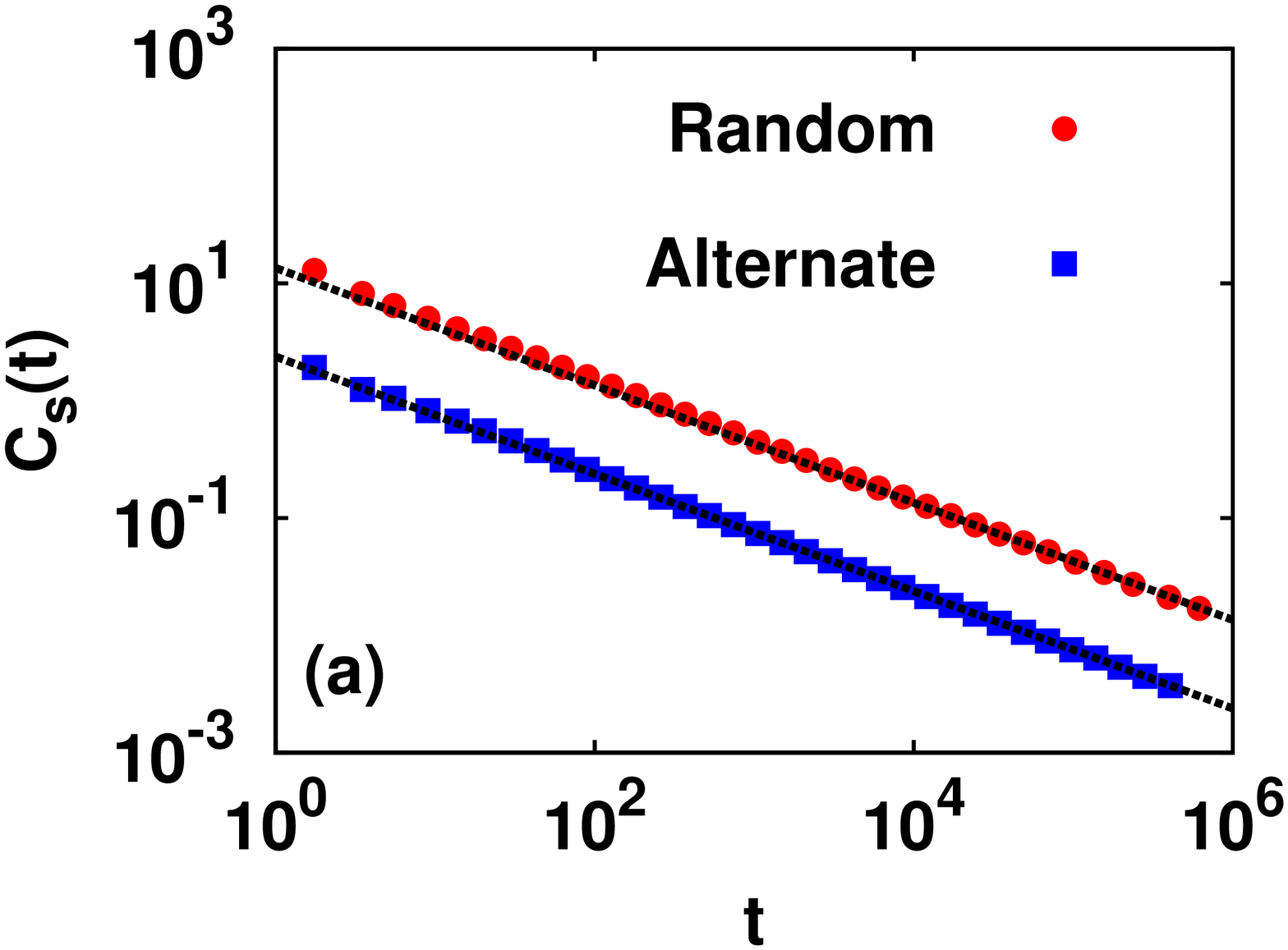}
\hskip-0.65cm
\includegraphics[width=3.5cm,angle=-90]{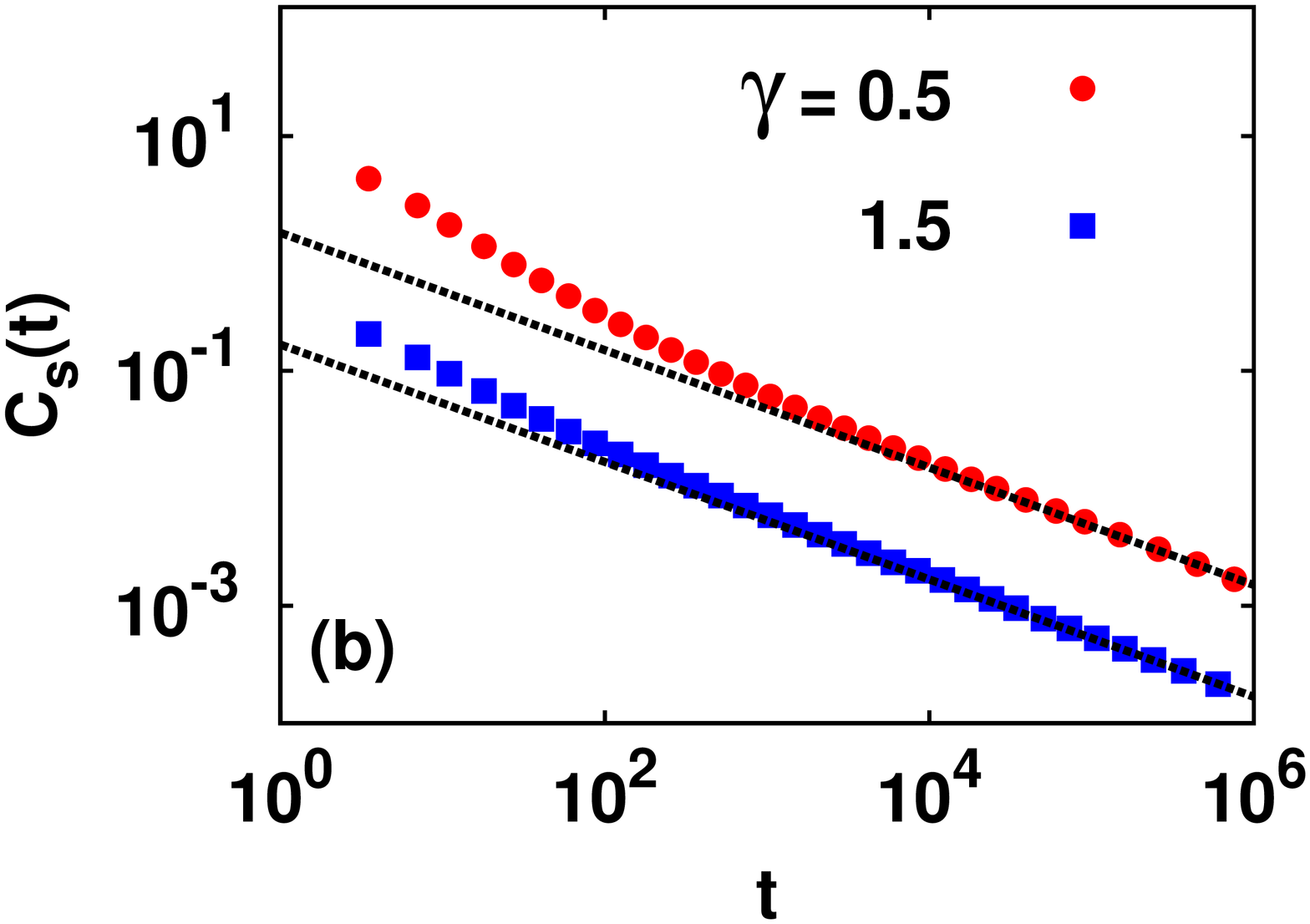}
}
\caption{
(Color online) 
Log-log plot of the autocorrelation function $C_s(t)$  
(a) for random coupling (circles) and alternate coupling
(squares) for $N = 20000$ with $\Delta t = 0.05$ 
(b) with anisotropy $\gamma = 0.5$ (circles), $1.5$ (squares)
for $N = 20000$ with $\Delta t = 0.10$.
The data is averaged over $100$  independent realisations and shifted along $y$-axis for clarity.
The broken lines have a slope $-0.5$.}
\label{fig:xxz}
\end{figure}

We have also simulated this model with other coupling schemes, namely, 
with alternate coupling $K_{i} = (-1)^i$, and random coupling ($K_i = \pm 1$ assigned randomly)
that have been studied in Ref. \cite{other}. The authors of Ref.  \cite{other} suggested that
spin diffusion appears to be diffusive for the alternate coupling case, whereas,
for random coupling, it is probably non-diffusive.
However, for both the cases we find a clear $t^{-1/2}$ behavior at late times.
This is shown in Fig. \ref{fig:xxz}a. Thus, our numerical results convincingly
demonstrate that in all the three models namely, uniform, alternate
and random coupling, the diffusion process is not anomalous.

We have studied spin diffusion in classical Heisenberg model
with anisotropic coupling in different spin directions
\begin{equation}
\mathcal{H}_{XXZ} = -K \sum_{i=1}^{N} \left[S_i^x  S_{i+1}^x + S_i^y  S_{i+1}^y + \gamma~S_i^z  S_{i+1}^z\right],
\end{equation}
where $\gamma$ is the anisotropy parameter. Spin diffusion with anisotropy
in the classical limit has been studied in some detail recently \cite{xxz-1, xxz-2, xxz-3}. 
Our data for $C_s(t)$ with $\gamma > 1$ and $\gamma < 1$ is shown in
Fig. \ref{fig:xxz}b.
% for $N = 20000$ and $\Delta t = 0.1$.
The data shows
that at late times $C_s(t) \sim t^{-1/2}$ and thus indicates normal spin
diffusion in the anisotropic model also.

\begin{figure}[]
\centerline
{
\includegraphics[width=4.10cm,angle=-90]{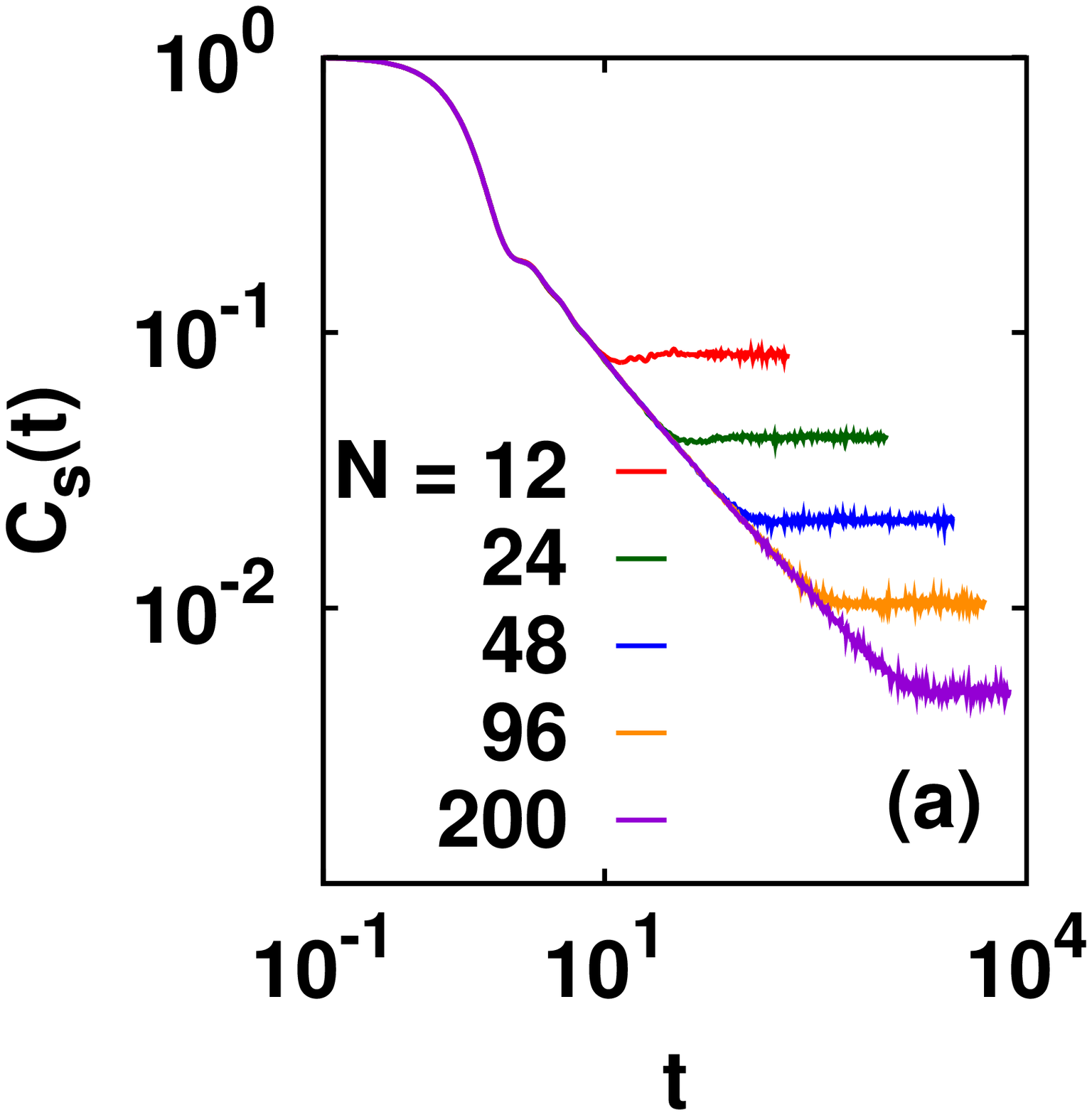}\hfill
\hskip-0.65cm
\includegraphics[width=4.10cm,angle=-90]{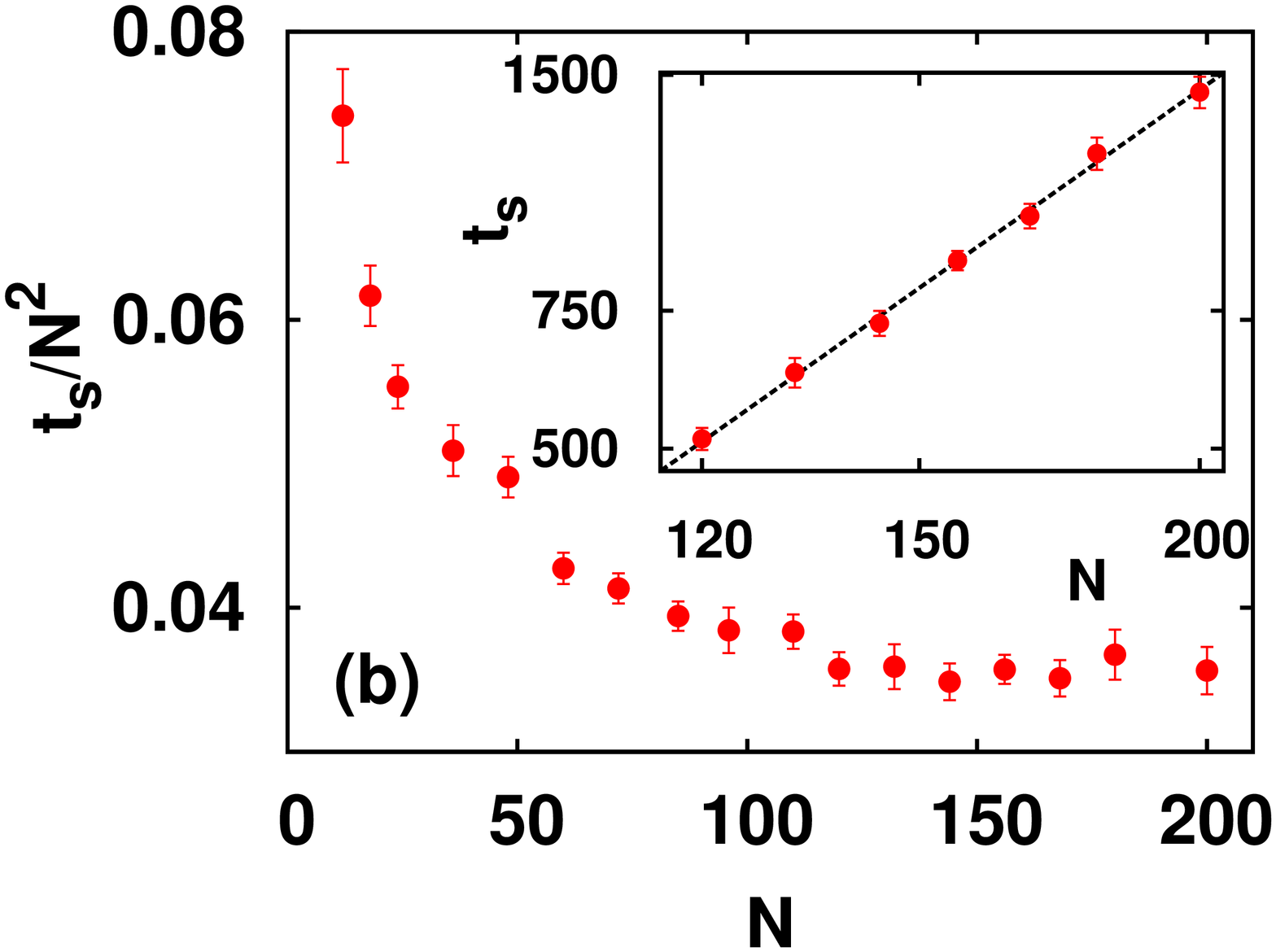}
}
\caption{(Color online) (a) $A_s(t)$ for finite systems $N$ saturate at different 
characteristic times $t_s(N)$.
(b) Plot of $t_s(N)/N^2$ as a function of $N$ which roughly becomes constant for $N > 100$.
The inset shows a log-log plot of $t_s(N)$ vs. $N$; the best fitted straight line for $N \geq 120$ 
gives a slope $z = 2.02 \pm 0.04$.}
\label{fig:sdev}
\end{figure}

Using finite size analysis of the ACF, one can have
an alternative method of estimating the exponent $\alpha$ \cite{nonlinear}.
We work with small system sizes $N \leq 200$ Heisenberg spins on a
ring with uniform coupling $K_i = 1$.
As stated earlier, $A_s(t)$ for
small system size saturates at some characteristic relaxation time
$t_s(N)$. The saturation of ACF for different $N$
is shown in Fig. \ref{fig:sdev}a.
Numerically, we estimate $t_s(N)$ by the intersection of a power
law fit, for the linear part of the
curve (in logarithmic scale) before it saturates, and the straight line
$y = A_s(\infty)$, where $A_s(\infty)$ is the
saturation value of $A_s(t)$ at late times. 
The saturation $A_s(\infty)$ is numerically computed by averaging 
$A_s(t)$ far away from the saturation point, i.e., for $t \gg t_s$.
The relaxation timescale $t_s(N)$ is related to the system size $N$
via the dynamical exponent $z$ as  $t_s \sim N^z$.
For a diffusive process $z$ should be equal to $2$, which is indeed the case as
can be seen from Fig. \ref{fig:sdev}b where we have plotted
$t_s/N^2$ against $N$; the data attains a constancy (approximately) for
$N > 100$. We fit the data for $N > 100$
which is shown in the inset of Fig. \ref{fig:sdev}b.
The best fitted straight line in the log-log plot has a 
slope $z = 2.02 \pm 0.04$. Hence, we
have $\alpha = z^{-1} \approx 0.5$, implying normal spin diffusion.
The value $z = 1.67$ reported earlier \cite{nonlinear}
seems to be due to the smaller system sizes studied there.

% 
% 
% \section{Conclusion}
% \label{conclusion}
To summarize, we have revisited the spin diffusion problem in
classical Heisenberg spin model in one dimension.
We have performed extensive simulation of the model using DTOE dynamics
that preserves the conservations of the total energy $E$ accurately.
Although this dynamics is identical to the equation of the motion only in
the $\Delta t \to 0$ limit, it equilibrates the system to the 
correct stationary state for any finite $\Delta t$ \cite{our}.
It is thus advantageous here to use a relatively larger $\Delta t$ 
and probe the dynamical behavior of the system up to late times. 
By computing the autocorrelation functions we show that, similar to energy diffusion,
spin diffusion in classical one dimensional Heisenberg model is normal ($\alpha = 1/2$), 
contrary to what has been suggested in some of the previous works.
We obtain an estimate for the dynamical exponent $z \approx 2$, which again
indicates that spin diffusion is normal.
The probable reasons as to why most of the previous works concluded that diffusion in this system
is anomalous could be because of (a) small scale simulations - both in system
size and time, (b) the noisy correlation function data, and (c) the accuracy of the method used.
Our way of simulation and analysis take care of most of these issues and
produce a clear long time diffusive tails for the correlation function.
Although this dynamics still lacks the strict total spin conservation, however
% spin conservation is appreciably maintained up to late times.
unlike conventional integration schemes where the errors accumulation is relatively fast,
here the total spin conservation is approximately preserved allowing one to measure the
autocorrelation functions up to very large time. It remains a challenge 
to find a suitable dynamics for this model, which will strictly preserves
both energy and spin conservation.

\vskip0.2cm
 
\textbf{Acknowledgement:} The author would like to thank P. K. Mohanty for 
stimulating discussions and careful reading of the manuscript.


\begin{thebibliography}{99}
\bibitem{fisher} M. E. Fisher, Am. J. Phys. \textbf{32}, 343 (1964)
\bibitem{joyce} G. S. Joyce, Phys. Rev. \textbf{155}, 478 (1967)


\bibitem{old} N. Bloembergen, Physica (Utrecht) \textbf{15}, 386 (1949)
\bibitem{hove} L. van Hove, Phys. Rev. \textbf{95}, 1374 (1954)


\bibitem{expt-1} D. Hone, C. Scherer, and F. Borsa, Phys. Rev. B \textbf{9}, 965 (1974).
\bibitem{expt-2} F. Borsa and M. Mali, Phys. Rev. B \textbf{9}, 2215 (1974).
\bibitem{expt-3} J-P. Boucher \textit{et al.} , Phys. Rev. B \textbf{13}, 4098 (1976).
\bibitem{expt-4} H. Benner, Phys. Rev. B \textbf{18}, 319 (1978);
\bibitem{expt-5} A. Lagendijk and E. Siegel, Solid State Commun. \textbf{29}, 709 (1976).

\bibitem{savin} A. V. Savin, G. P. Tsironis, and X. Zotos, Phys. Rev. B \textbf{72}, 140402(R) (2005)
\bibitem{our} D. Bagchi and P. K. Mohanty,  arXiv:1206.2827 (to appear in Phys. Rev. B, 2012)
\bibitem{goldstein} H. Goldstein, C. P. Poole and J. L. Safko, Classical Mechanics, 3rd Edition, Addison Wesley.

%Thermally driven classical Heisenberg model in one dimension,

\bibitem{muller} G. M\"{u}ller, Phys. Rev. Lett. \textbf{60}, 2787 (1988); \textbf{63}, 813 (1989)
\bibitem{landau} 
R. W. Gerling and D. P. Landau, Phys. Rev. Lett. \textbf{63}, 812 (1989); Phys. Rev. B \textbf{42}, 8214 (1990).

\bibitem{error} 
J. Liu, N. Srivastava, V. S. Viswanath, and G. M\"{u}ller, J. Appl. Phys. \textbf{70}, 6181 (1991)

\bibitem{breakdown}
O. F. de Alcantara Bonfim and G. Reiter, Phys. Rev. Lett. \textbf{69}, 367 (1992) 
\bibitem{breakdown-comment}
M. B\"{o}hm, R. W. Gerling, and H. Leschke, Phys. Rev. Lett. \textbf{70}, 248 (1993)

\bibitem{mct} S. W. Lovesey and E. Balcar, J. Phys.: Condens. Matter \textbf{6} (1994);
S. W. Lovesey, E. Engdabl, A. Cuccoli, V. Tognetti and E. Balcar, J. Phys.: Condens. Matter \textbf{6} (1994)

\bibitem{other}
N. Srivastava, J. Liu, V. S. Viswanath, and G. M\"{u}ller, J. Appl. Phys. \textbf{75}, 6751 (1994)

\bibitem{nonlinear} V. Constantoudis and N. Theodorakopoulos, Phys. Rev. E \textbf{55}, 7612 (1997)

\bibitem{xxz-1} S. Davis, and G. Guti\'{e}rrez, Physica B \textbf{355}, 1 (2005)
\bibitem{xxz-2} D. L. Huber, Physica B \textbf{407}, 4274 (2012)
\bibitem{xxz-3} R. Steinigeweg, EPL, \textbf{97} 67001 (2012) 


\end{thebibliography}
\end{document}